\documentclass[noshowpacs,amsmath,
twocolumn,
superscriptaddress,
8pt,
aps,prb]{revtex4-1}
\usepackage{setspace}
\usepackage{amsmath}
\usepackage{graphicx}
\usepackage[nearskip,margin = 0pt]{subfig}

\usepackage{verbatim}
\usepackage{amsfonts}
\usepackage{amssymb}
\usepackage{epstopdf} 
\usepackage{xcolor}
\usepackage{verbatim}
\usepackage{upgreek}
\usepackage{ragged2e}
\usepackage{url}
\usepackage[hidelinks]{hyperref}
\DeclareGraphicsExtensions{.pdf,.eps,.png,.jpg,.mps}

\begin{document}

\title{Heterogeneously Integrated Squeezed-Light Generation and Detection on a Single Photonic Chip}














\author{Haoran Chen$^{1}$, Benjamin Westcott$^{1}$, Fatemehsadat Tabatabaei$^{1}$, Xiangwen Guo$^{1}$, Shuman Sun$^{1}$, Zijiao Yang$^{1,2}$, Gedalia Y. Koehler$^{1}$, Beichen Wang$^{1}$, Shadrach Sarpong$^{1}$, Steven Bowers$^{1}$, Olivier Pfister$^{2}$, Andreas Beling$^{1,\dagger}$ and Xu Yi$^{1,2,\dagger}$\\
\vspace{3pt}
$^1$Department of Electrical and Computer Engineering, University of Virginia, Charlottesville, Virginia 22904, USA.\\
$^2$Department of Physics, University of Virginia, Charlottesville, Virginia 22904, USA.\\
$^{\dagger}$Corresponding authors: andreas@virginia.edu, yi@virginia.edu}

\begin{abstract}
\noindent Squeezed light underpins quantum-enhanced sensing and continuous-variable quantum information processing, and integrated photonics offers a route to producing it at scale. Universal to these applications are squeezed-light generation and measurement. Importantly, quantum measurements serve not only as readout but also as active operations in quantum-state evolution. However, integrating squeezed-light generation and photodetection on the same photonic chip has remained challenging because they impose fundamentally conflicting material requirements: low optical loss to preserve quantum correlations, but efficient photon absorption for photodetection. Here, we demonstrate squeezed-light generation, routing, and balanced homodyne detection integrated on a single photonic chip through heterogeneous integration. A two-mode squeezed quantum micro- comb comprising 34 quantum modes is measured with approximately 3 dB squeezing. Our work establishes a scalable architecture for fully integrated squeezed-light quantum photonic systems, unifying quantum-state generation, processing, and detection on a single chip.

\end{abstract}
\date{\today}

\maketitle

\noindent {\bf Introduction}










Squeezed light exploits quantum correlations to suppress fluctuations in one quantum field quadrature below the shot-noise limit at the expense of increased fluctuations in the conjugate quadrature\cite{andersen201630}. It has transformed precision sensing by enabling quantum-enhanced measurements, most notably in gravitational-wave detection with LIGO\cite{tse2019quantum} and nonlinear spectroscopy of biological molecules\cite{casacio2021quantum}. Beyond sensing, multimode squeezed states constitute a fundamental resource for continuous-variable quantum information processing\cite{braunstein2005quantum} and measurement-based quantum computing (MBQC)\cite{Raussendorf2001,menicucci2006universal,briegel2009measurement,menicucci2014fault,chen2014experimental,Yoshikawa2016,asavanant2019generation,larsen2019deterministic}. These applications require the generation and measurement of thousands—or ultimately millions—of entangled quantum modes (qumodes), placing stringent demands on scalability, optical loss, and circuit complexity. This has motivated intensive efforts to develop integrated photonic platforms for squeezed-light generation, leading to demonstrations of single-mode squeezing \cite{zhao2020near,zhang2021squeezed,chen2022ultra,park2024single,ulanov2025quadrature}, two-mode squeezing\cite{vaidya2020broadband,liu2025wafer}, intensity squeezing\cite{dutt2015chip,shen2025highly}, squeezed microcombs\cite{yang2021squeezed,jahanbozorgi2023generation,wang2025large,jia2025continuous,li2026spectrally,shen2025highly}, and ultrafast squeezed light\cite{nehra2022few}.

\medskip


In parallel, significant progress has also been made toward the integrated measurement of squeezed light\cite{raffaelli2018homodyne, tasker2021silicon}.
Unlike classical systems, however, quantum measurements do not merely extract information but actively participate in quantum state evolution. For two-mode and multimode squeezed light, measurements project the quantum state of the remaining modes, enabling key protocols including quantum teleportation \cite{furusawa1998unconditional}, Gottesman–Kitaev–Preskill (GKP) state generation\cite{vasconcelos2010all,Eaton2022_PhANTM,konno2024logical,larsen2025integrated}, and more generally, measurement-based quantum computing\cite{raussendorf2003measurement,briegel2009measurement,menicucci2014fault}. As a result, scalable quantum photonic systems require not only integrated quantum light sources but also equally scalable quantum measurements. Bringing these functionalities onto the same chip is essential to preserve optical coherence, minimize loss, and enabling low-latency feedforward required for large-scale quantum information processing.



\begin{figure*}[!bht]
\captionsetup{singlelinecheck=off, justification = RaggedRight}
\includegraphics[width=17cm]{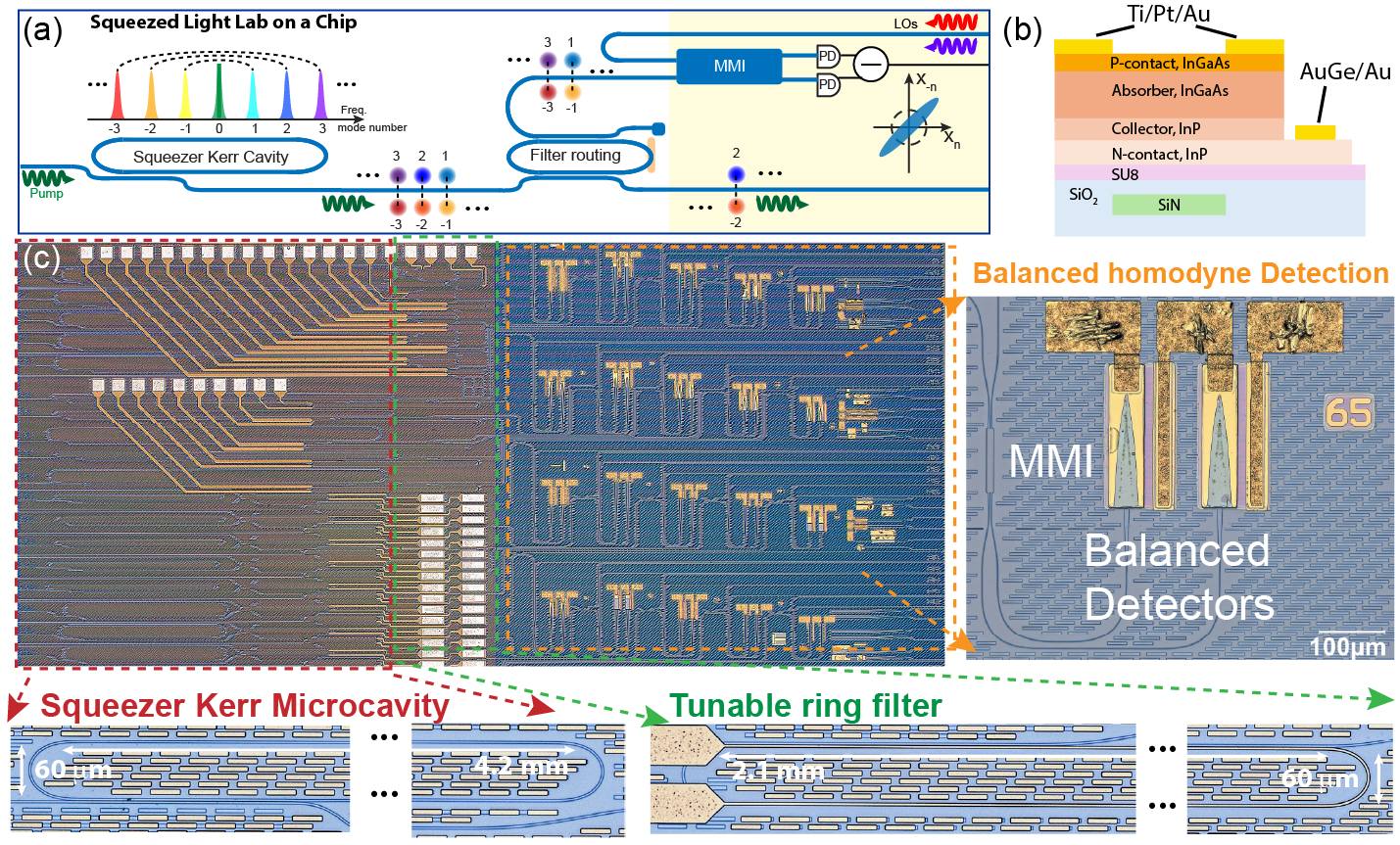}
\caption{{\bf Architecture and building blocks of the heterogeneous integrated squeezed-light generation and detection on a photonic chip.} 
\textbf{(a)} Illustration of the heterogeneous quantum photonic integrated circuit. The quantum light source is a squeezed quantum microcomb generated in a SiN optical microcavity through Kerr nonlinearity. The squeezer is followed by a tunable ring filter to route the odd-number quantum modes (qumodes) to the balanced homodyne detection (BHD). The routed odd number qumodes are then combined with local oscillator on a 50/50 multimode interferometry (MMI) coupler, and detected on a pair of balanced MUTC photodiodes heterogeneously integrated on the SiN QPIC. \textbf{(b)} Sideview illustration of the heterogeneously integrated MUTC photodiodes. \textbf{(c)} Microscopic images of the heterogeneously integrated photonic chip with many paralleled squeezed light generation and detection circuits. Typical zoom-in images of the squeezer, the tunable filter, and the balanced photodiodes are also shown.}
\label{fig:picture}
\end{figure*}

\medskip

Despite substantial progress in integrated squeezed-light sources and quantum receivers, squeezed-light generation and photodetection, e.g., balanced homodyne detection, have not been brought together on the same photonic chip and have remained physically separated\cite{jia2026monolithic}.
The fundamental challenge lies in the conflicting requirements of quantum state generation and measurement: squeezed-light generation and routing require ultra-low-loss photonic circuits to preserve fragile quantum correlations, whereas measurement requires complete optical absorption in high-efficiency photodetection. These conflicting requirements are near impossible to satisfy within a single material platform. Heterogeneous integration offers a scalable route to combine low-loss quantum photonic circuits with integrated photodetectors, enabling fully integrated squeezed-light quantum photonic systems.


\medskip

Here, we demonstrate squeezed light generation, routing, and  balanced homodyne detection integrated within a single SiN photonic chip. The quantum photonic integrated circuit (QPIC) combines a high-Q Kerr microcavity for two-mode vacuum squeezed quantum microcomb generation \cite{jahanbozorgi2023generation}, an FSR-engineered tunable ring filter for qumode routing, and a balanced homodyne receiver consisting of a 50/50 multimode interference (MMI) coupler and a pair of heterogeneously integrated modified uni-traveling carrier (MUTC) photodiodes fabricated through wafer bonding \cite{yu2020heterogeneous,gao2025heterogeneous}. As a result, squeezed-light generation, routing, and balanced homodyne detection (BHD) are all performed on the same chip, and the squeezed states never leave the integrated photonic chip. We measure approximately 3 dB of raw squeezing and demonstrate 34 qumodes in the form of 17 pairs of two-mode squeezing. The post-squeezer circuit, including the filter, MMI coupler, and photodiodes, has a total optical loss of 1.1 dB, highlighting the low-loss nature of the integrated measurement circuit. The demonstrated architecture provides a scalable path towards fully integrated squeezed light quantum photonics systems, and can be widely useful for quantum sensing, continuous-variable quantum information processing, and measurement-based photonic quantum computing.

\medskip

\noindent {\bf Results.}






\begin{figure*}[!bht]
\captionsetup{singlelinecheck=off, justification = RaggedRight}
\includegraphics[width=17cm]{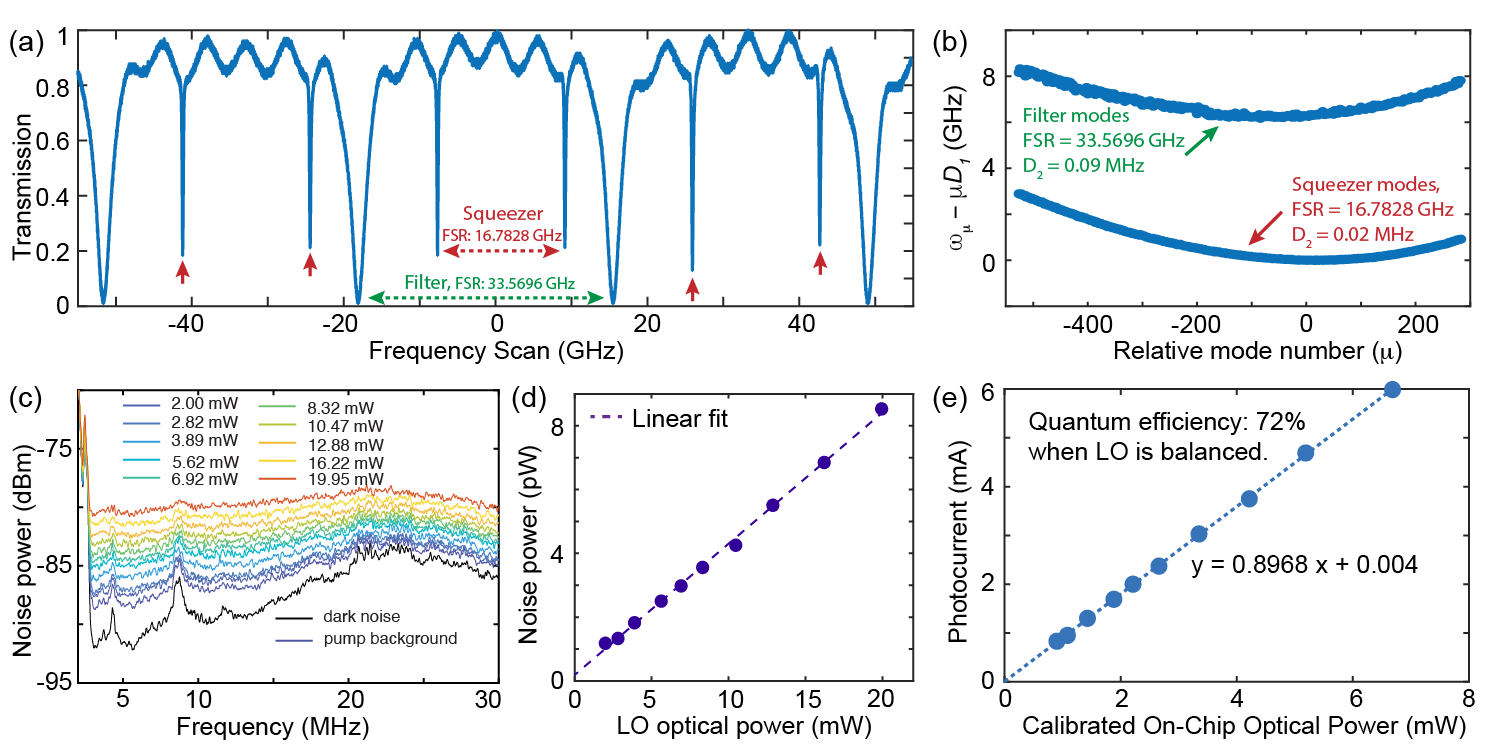}
\caption{{\bf Characterization of components in the squeezed-light generation and detection chip.}
\textbf{(a)} Transmission spectrum of the squeezer and the tunable filter. The filter frequencies can be tuned by an integrated heater to align with the resonance frequencies of the squeezer. \textbf{(b)} Mode spectrum of the squeezer cavity and the tunable filter cavity. The FSR of the filter cavity matches twice of the squeezer cavity FSR within 4 MHz difference. \textbf{(c)} Noise spectrum of the BHD at different LO optical power levels. \textbf{(d)} The noise power increase linearly with the LO optical power, validating the BHD is shot noise limited. \textbf{(e)} Total quantum efficiency measurement of the filter, MMI coupler and balanced photodiodes. Measurement is performed when the bias voltages of the PDs are adjusted to balance the LO to shot noise limit.}
\label{fig:device}
\end{figure*}

\begin{figure*}[!bht]
\captionsetup{singlelinecheck=off, justification = RaggedRight}
\includegraphics[width=17cm]{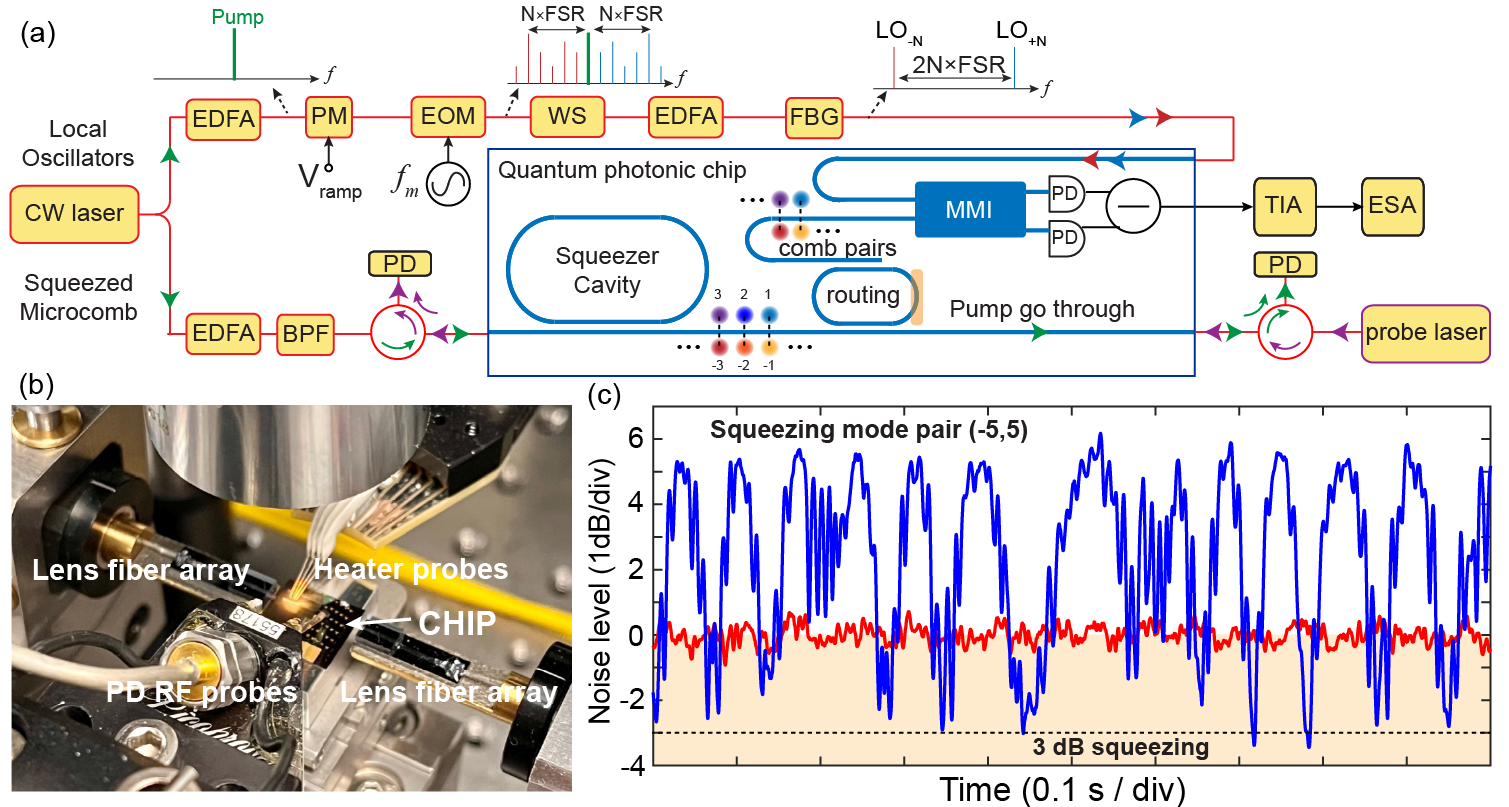}
\caption{{\bf Experimental setup and squeezing measurement.}
\textbf{(a)} Experimental setup. The pump laser and the local oscillators are created off-chip and coupled into the squeezed light chip through lens fiber array. The bichromatic LOs are created using electro-optics modulation and a line-by-line programmable waveshaper (WS) filter. A probe laser is coupled into the chip from the opposite direction to monitor the frequency alignment of the squeezer and the tunable filter. A low bandwidth DC probe is used to control the integrated heater on the ring filter, and an RF probe is used to bias the integrated photodiodes and read-out the photocurrents. \textbf{(b)} Image of the squeezed light generation and measurement chip setup. \textbf{(c)} Two-mode squeezing measurements of qumode pair (-5,5). Representative quadrature noise variance (blue) relative to shot noise (red) as a function of time while the phase of the LOs are ramped to yield squeezing and anti-squeezing periodically. A horizontal dash line indicates 3-dB squeezing level. Photocurrent background from the dark noise and the leaked pump are subtracted for both quadrature noise variance and the shot noise traces. The orange shaded box highlights the area below shot noise level. The measurements are taken at 2.7 MHz frequency, 100 kHz resolution bandwidth, and 100 Hz video bandwidth.}
\label{fig:3dB}
\end{figure*}

\begin{figure*}[!bht]
\captionsetup{singlelinecheck=off, justification = RaggedRight}
\includegraphics[width=17cm]{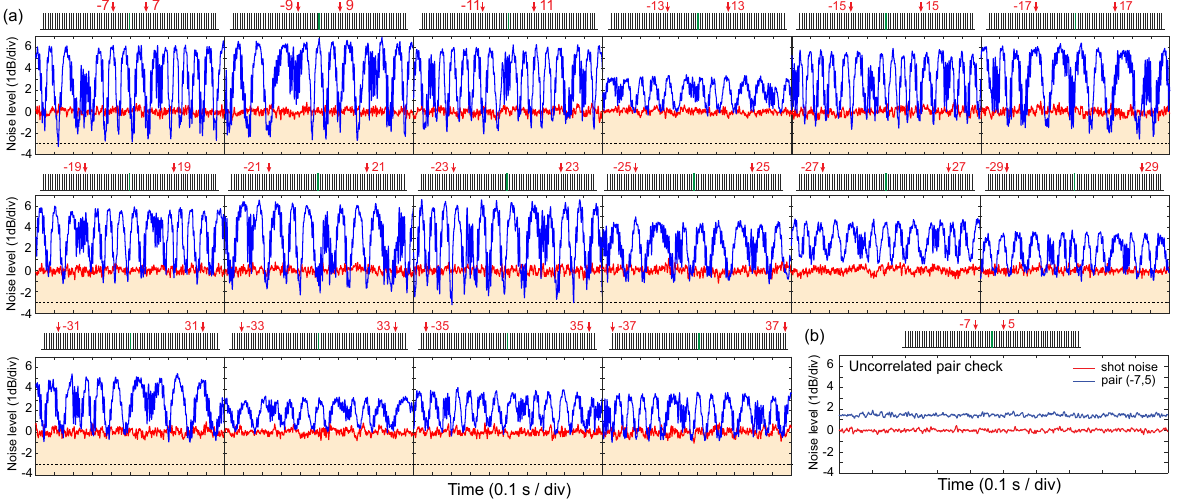}
\caption{{\bf Squeezed quantum microcomb generation and measurement on the heterogeneously integrated chip.}
\textbf{(a)} Quadrature noise variances (blue) relative to shot noise (red) of odd number qumodes from (-7,7) to (-37,37). The qumodes measured are marked by the red arrows. The regime below the shot noise limit is colored in orange, and a dashed black line indicates 3 dB below the shot noise level. Qumodes are selected in measurements by tuning the frequency of the LOs to align with the target qumodes. \textbf{(b)} Quantum correlation check: noise variances show no quantum correlation between uncorrelated comb pairs for qumodes (-7,5). }
\label{fig:comb}
\end{figure*}

The quantum photonic integrated circuit (QPIC) is based on low-loss 800-nm thick SiN photonic platform. In the QPIC (Fig.\ref{fig:picture}a), a high-Q microresonator is used to generate the squeezed quantum microcomb\cite{jahanbozorgi2023generation}, and the odd-number comb lines are routed to the drop port of a tunable ring filter. The squeezed light is then combined with the local oscillator (LO) on a multi-mode interferometry (MMI) coupler and sent to a pair of heterogeneously integrated balanced photodiodes \cite{yu2020heterogeneous} for balanced homodyne detection (BHD). In this configuration, the squeezed light never leaves the chip: it is generated, routed and detected on the same chip. The classical light used in the system, e.g., pump laser and LOs, are generated off-chip and coupled into the chip through lensed fiber arrays. Microscopic pictures of the QPIC and its components are shown in Fig.\ref{fig:picture}.


The squeezer is a waveguide-coupled racetrack-shaped ring resonator with a cross-section of 800 nm and $1.55$ $\mu$m. It has a free-spectral-range (FSR) of 16.78 GHz, a loaded (intrinsic) Q-factor of 0.89 (4.35) million, and an escape efficiency around 75\% at 1550 nm. The on-chip OPO threshold of the squeezer is 245 mW. The filter is a racetrack-shaped resonator with symmetric over-coupling through port and drop port. The filter is near critical coupling at the through port, with a loaded Q of 0.18 million, corresponding to a full-width half maximum (FWHM) of 1.07 GHz. The FSR of the filter is designed to be twice of that of the squeezer, so that the odd-number comb lines can be picked up by the drop port, while the strong pump light transmits to the through port. The FSR of the filter is measured to be 4 MHz away from 2$\times$FSR of the squeezer, and this deviation is much smaller than the FWHM of the filter (Fig. \ref{fig:device}a,b). When the filter is aligned with the squeezed modes in frquency, the isolation ratio of the pump light at the through port is measured to be around $27$ dB, which is compromised by a nearby resonance mode of the other polarization. The MMI splitting ratio is measured to be 46/54, while it was designed to be 50/50. As a result, the bias voltage of a photodiode is adjusted to lower its quantum efficiency (QE) by 16\%, so that the LOs are balanced in the BHD to reach shot noise limit despite the unbalanced MMI coupler. 

The photodiodes used for balanced detection are InAlGaAs/InP modified uni-traveling carriers (MUTC). 
Fig. \ref{fig:picture}b shows a simplified cross-section of the heterogeneous PD. Our photodiodes feature a thick (1$\mathrm{\mu m}$) absorber and a 0.2 $\mathrm{\mu m}$ collector to reduce capacitance. Tight integration was achieved using die-to-wafer adhesive bonding with 63 nm-thick SU8 as the bonding agent. The epilayer stack was bonded P-side up, placing the N-contact directly above the waveguide. Ti/Pt/Au and AuGe/Au metallizations were used for the p-contact and n-contact respectively, and radio frequency (RF) pads and their connecting air-bridges were formed by gold (Au) plating and connect the two PDs in an anti-parallel fashion. We have shown previously that Ti/Pt/Au p-contact metallizations can lead to significant absorption loss in evanescently coupled waveguide photodiodes due to the high conductivity of Titanium and Platinum around 1.55 $\mathrm{\mu m}$ \cite{ben2026cleo}. This means that p-metal area and quantum efficiency are inversely related for this particular metallization. In order to obtain high quantum efficiencies without compromising factors which scale alongside p-metal area, most notably depletion voltage, the p-metal profile was designed such that its spatial overlap with the waveguide + PD mode is minimized. This leads the taper shape shown in Fig. \ref{fig:picture}c. Each PD has an area of 225x60 $\mathrm{\mu m}^{2}$, of which, 73.2\% is covered in p-metal. The average dark current for our taper design is 0.11 $\mathrm{\mu A}$ at -3V bias voltage.
The non-destructive loss measurement method \cite{chen2026universal} is used to characterize the facet coupling loss and the quantum efficiency of the QPIC. The left coupling facet loss is 1.5 dB, while the on-chip total quantum efficiency of the filter, MMI and photodiodes, is measured to be 72\%, with the PDs bias set for balancing LOs on the BHD. The total QE of the circuit would be 78\% if the one of the PD's QE is not deliberately tuned down for LO balancing.

The experimental configuration is shown in Fig. \ref{fig:3dB}. A continuous-wave (cw) laser is amplified to pump the squeezer. The on-chip pump power is set to 234 mW, which is 0.2 dB below the OPO threshold. Part of the cw laser is split to derive bichromatic LOs through electro-optics modulation and line-by-line waveshaping, which is described in detail elsewhere \cite{yang2021squeezed}. The LOs can be tuned to different frequencies to measure different quantum modes. To align the frequency of the filter and the squeezer mode, a tunable auxiliary cw laser is coupled from the right side of the chip, and its frequency is ramped to yield the transmission of the squeezer cavity and the filter cavity. The filter frequency can be adjusted by tuning the integrated heater. As the auxiliary laser is propagating in the opposite direction of the pump and the squeezed light, it does not interfere with the squeezing generation and detection. It can also be switched off once the filter and the squeezer cavity are aligned in frequency. 

The quadrature noise variances of 17 sets of comb pairs (34 qumodes) are measured using the integrated balanced homodyne detection. A typical two-mode squeezing measurement result is shown for mode pair (-5,5) in Fig. \ref{fig:3dB}c, where the quadrture noise variance is shown in blue and the shot noise is shown in red. A dash line indicates 3 dB below shot noise level. In the measurement, the phase of the LOs is ramped to yield varying quadrature variances. The measurements are taken with an electrical spectrum analyzer in zero-span mode, at 3.5 MHz offset frequency, 100 kHz resolution bandwidth, and 100 Hz video bandwidth. The noise floor contributed by the leaked pump and dark current is subtracted for both the shot noise and the quadrature noise variance in both Fig. \ref{fig:3dB} and Fig. \ref{fig:comb}. The noise variance of mode pairs (-7,7) to (-37, 37) are presented in Fig. \ref{fig:comb}. The reduced anti-squeezing and squeezing level at large mode number could be a result of the resonance frequency walk-off between the squeezer and the filter, due to the small mismatch of their FSR. Further investigation is needed to fully understand the squeezing and anti-squeezing variation versus mode number.

\medskip

A squeezing level of 3 dB for mode (-5,5) is close to the theory prediction for our QPIC, and it is achieved with a QPIC far from perfect. For the resonator with 75\% escape efficiency and dispersion of $D_2 = 20$ kHz, the analytical calculation suggests 5.4 dB of squeezing in the waveguide when the pump power is 95 \% of the OPO threshold. Given the total quantum efficiency of 72\% after the squeezer (8\% degradation from the unbalanced MMI), the measured squeezing will be reduced to 3.1 dB which is close to our measured values. Increasing the escape efficiency and optimizing the MMI splitting ratio could dramatically increase the squeezing level. For 90\% escape efficiency and 80\%  quantum efficiency after the squeezer, the expected measured squeezing can reach $4.9$ dB. This is a very reasonable near-term goal. When comparing the current results to our own SiN squeezing measurements with off-chip detection\cite{jahanbozorgi2023generation}, we have improved the squeezing from 1.1 dB to 3 dB, despite the escape efficiency of the squeezer in this work (75\%) is much lower than the previous work (89\%). The improvement is mainly because the total quantum efficiency after the squeezer has increased from 35\% to 72\%, by eliminating the 2.3 dB waveguide to fiber coupling loss, and the 2 dB off-chip filter loss. This showcases the advantage of integrating all squeezed light components on a single chip.
Another factor that could limit our squeezing measurement is the relative phase noise between the squeezer and the LO path. Phase jitter is clearly presented in our noise variance traces, and phase stabilization of the LO path relative to the squeezed light path will certainly reduce the impact of phase noise.


\medskip
\noindent {\bf Discussion.}
In summary, we have demonstrated, to our knowledge, the first integration of squeezed-light generation and balanced homodyne detection on a heterogeneously integrated photonic chip. The demonstrated 3 dB squeezing is currently limited by component-level non-idealities, including the escape efficiency of the squeezer, the splitting imbalance of the MMI coupler, and residual circuit loss, and can be further improved through optimization of individual components and system-level integration. Recently, waveguide-based optical parametric amplifiers have demonstrated squeezing levels approaching 10 dB \cite{kashiwazaki202610}, suggesting that such performance can ultimately be translated to integrated photonic platforms.
Beyond balanced homodyne detection, alternative approaches for squeezed-light characterization, e.g., optical homodyne detection schemes \cite{shaked2018lifting,nehra2022few}, provide additional opportunities for scalable quantum measurements. Looking forward, the heterogeneous integration approach demonstrated here can be extended to incorporate single-photon detectors and photon-number-resolving detectors \cite{cheng2023100}, enabling non-Gaussian operations and hybrid quantum functionalities directly on chip. We envision that the integration of quantum state generation, low-loss photonic processing, and advanced quantum detection within a unified photonic platform will provide a scalable foundation for future continuous-variable quantum computing, quantum sensing, and quantum networking.

\medskip

\noindent\textbf{Methods}
\begin{footnotesize}

\noindent{\bf Shot noise calibration.}
In our experiment, the shot noise is measured by tuning the LOs frequency 500 MHz away from the squeezer frequency, and measure the noise variance of the vacuum there. The reason we took this approach is because a small amount of pump laser is leaked into the BHD, and it is not balanced on the BHD due to the unbalanced MMI splitting ratio. As a result, the leaked pump is contributing an non-neglected level of noise floor (Fig. \ref{fig:device}c, 2nd last trace). While it can be treated the same way as photodiode dark noise, the noise floor introduced by the leaked pump does depend on whether the pump laser is on resonance with the squeezer or not. Therefore, the shot noise has to be measured when the pump is on resonance to ensure that the same amount of leaked pump power is on the BHD for both the squeezing noise variance and the shot noise measurements. In our earlier work, we have verified that tuning the LO frequency away from the squeezing frequency will yield the measured noise variance to converge to the shot noise \cite{yang2021squeezed}. We verified the measurements are shot noise limited as the noise power increases linearly with the LO power (Fig. \ref{fig:device}d), with or without the leaked pump into BHD.

\noindent{\bf MMI splitting ratio measurement.} In the measurement, laser light is first sent into the MMI through the LO port and photocurrents on both photodiodes are recorded ($I_{1}$, $I_{2}$). The photocurrent ratio only relies on the MMI splitting ratio and the quantum efficiency ratio of the two detectors: $I_1/I_2 = T\eta_1 / R \eta_2$. Here we assumed the MMI to be unitary, such that the transmission from port 1 to 3 equals that from port 2 to 4, and is defined as $T$, and transmission from port 1 to 4 equals that from port 2 to 3, and is defined as $R$. The photodiodes quantum efficiencies are defined as $\eta_1, \eta_2$. Then the laser light is sent into the MMI through the filter port, and the photocurrents on the detectors are recorded again ($I'_{1}$, $I'_{2}$), and $I'_1/I'_2 = R\eta_1/T\eta_2$. As a result, the splitting ratio can be expressed as $T/R = \sqrt{I_1 I'_2/I_2 I'_1}$, and the photodiodes quantum efficiency ratio is $\eta_1/\eta_2 = \sqrt{I_1 I'_1/I_2 I'_2}$. We measured $T/R = 0.862$, and $\eta_1/\eta_2 = 1.005$ when both of the PDs are biased for optimal quantum efficiency. 




\end{footnotesize}

\medskip

{\noindent \bf Note}—During the preparation of this manuscript, we became aware of related independent work reporting on-chip generation and detection of squeezed light in a silicon photonic platform \cite{green2026generation}.

\noindent\textbf{Acknowledgement}

\noindent The authors acknowledge Ligentec for SiN photonic circuit fabrication, and gratefully acknowledge DARPA INSPIRED (HR0011-24-2-0360), National Science Foundation (1842641, 2238096, 2531569), DOE (DE-SC0023337) and QC82 Inc. O.P. was supported by National Science Foundation PHY-2514971, and ECCS-2530171. The views and conclusions contained in this document are those of the authors and should not be interpreted as representing official policies of DARPA, DOE, or the U.S. Government.

\medskip

\noindent \textbf{Author Contributions}\\ 
X.Y., A.B., O.P. conceived the concept of the experiments. H.C. performed the measurements with assistant from G.K., S.S.. B.W., F.T., X.G. designed and fabricated the integrated photodiodes. S.S., Z.Y., B.W., H.C., designed the photonic integrated circuits. H.C., X.Y., O.P., A.B. analyzed the experimental results. X.Y., A.B. supervised the experiments. All authors participated in preparing the manuscript.

\medskip

{\noindent \bf Competing interests}
The authors declare no competing interests.

\medskip

{\noindent \bf Data availability.} The data that support the plots within this paper and other findings of this study are available from the corresponding authors upon reasonable request.

\medskip

{\noindent \bf Code availability.} The codes that support the findings of this study are available from the corresponding authors upon reasonable request.

\bibliographystyle{naturemag}
\bibliography{ref}

\end{document}